\documentclass[twoside,onecolumn,aps,prc,superscriptaddress,
nofootinbib,showpacs]{revtex4}
\usepackage{amsmath,amssymb,graphicx}

\usepackage{multirow}
\usepackage{amsfonts}
\usepackage{dsfont}
\usepackage{subfigure}
\usepackage{amssymb}
\usepackage{amsmath}
\usepackage{amsfonts}
\usepackage{epsfig}
\usepackage{graphicx}
\usepackage{graphics}
\usepackage{subfigure}
\usepackage{verbatim}

\usepackage{color}
\makeatletter

\newcommand{\Rmnum}[1]{\expandafter\@slowromancap\romannumeral #1@}
\makeatother

\allowdisplaybreaks

\begin{document}

\title{Possible $D\bar{D}$ and $B\bar{B}$ Molecular states in a chiral
quark model}

\author{M.T. Li}
\affiliation{Institute of High Energy Physics, CAS, P.O. Box 918-4,
Beijing 100049, China}
\affiliation{Theoretical Physics Center for Science Facilities (TPCSF),
CAS, Beijing 100049, China}

\author{W.L. Wang}
\affiliation{Institute of High Energy Physics, CAS, P.O. Box 918-4,
Beijing 100049, China}
\affiliation{Theoretical Physics Center for Science Facilities (TPCSF),
CAS, Beijing 100049, China}

\author{Y.B. Dong}
\affiliation{Institute of High Energy Physics, CAS, P.O. Box 918-4,
Beijing 100049, China}
\affiliation{Theoretical Physics Center for Science Facilities (TPCSF),
CAS, Beijing 100049, China}

\author{Z.Y. Zhang}
\affiliation{Institute of High Energy Physics, CAS, P.O. Box 918-4,
Beijing 100049, China}
\affiliation{Theoretical Physics Center for Science Facilities (TPCSF),
CAS, Beijing 100049, China}

\begin{abstract}

We perform a systematic study of the bound state problem of $D\bar{D}$
and $B\bar{B}$ systems by using effective interaction in our chiral
quark model. Our results show that both the interactions of $D\bar{D}$
and $B\bar{B}$ states are attractive, which consequently result
in $I^G(J^{PC})=0^+(0^{++})$ $D\bar{D}$ and $B\bar{B}$  bound states.

\end{abstract}



\maketitle

\section{Introduction}

Since the discovery of X(3872), many X, Y, and Z exotic states have
been reported. These hadrons have ever been explained as molecules,
tetraquarks, hybrids {\it et al.} because they can't be interpreted
as simple quarkoniums. In our previous chiral quark model
calculation \cite{li1}, it is found that the newly observed hadrons,
such as $Z_b(10610)$, $Z_b(10650)$, $X(3872)$ and $Y(3940)$, might
be assigned as the $B\bar{B}^*$, $B^*\bar{B}^*$, $D\bar{D}^*$ and
$D^*\bar{D}^*$ bound states. These results stimulate our further
interest in studying their analogues, i.e. the systems of $D\bar{D}$
and $B\bar{B}$.

So far, several works have been done to calculate the $D\bar{D}$ and
$B\bar{B}$ states
\cite{wong,huang,valcarce1,valcarce2,valcarce3,liux1,
liux2,liux3,ke,yang}. Valcarce et al.
\cite{valcarce1,valcarce2,valcarce3} considered the $D\bar{D}$
coupled to charmonium-light two-meson systems, like $J/\psi \omega$
channel, and concluded that the $0^+(0^{++})$ $D\bar{D}$ is the only
possible bound state.  Ke {\it et al.} \cite{ke} supported the
existence of $0^+(0^{++})$ $D\bar{D}$ and $B\bar{B}$ molecules in
the Bethe-Salpeter framework. In one-meson-exchange model, Liu et
al. \cite{liux1,liux2,liux3} calculated the binding energies of
$D\bar{D}$ and $B\bar{B}$ systems. Meanwhile, Zhang {\it et al.}
\cite{huang} got such molecules: $B\bar{B}$ of 10580$\pm$100 MeV and
$D\bar{D}$ of 3760$\pm$100 MeV in QCD sum rule calculation on the
quark level. Their results are consistent with Wong's prediction in
a two-gluon-exchange model \cite{wong}. On the contrary, Yang {\it
et al.} \cite{yang} argued that in color-singlet channel $D\bar{D}$
and $B\bar{B}$ molecules didn't exist. To sum up the above
calculations, one sees that the existence and properties of the
possible $D\bar{D}$ and $B\bar{B}$ molecular states are presently
model dependent. Further theoretical investigations are expected to
be significant.

In this paper, we perform a dynamical study of the $D\bar{D}$ and
$B\bar{B}$ systems with isospin $I=0$ and $1$ in our chiral quark
model by using the effective interaction. The chiral quark model was
built in such a way that the chiral symmetry is  restored by
introducing the coupling between quark field and Goldstone bosons
and the constituent quark mass is obtained as a consequence of the
spontaneous vacuum symmetry breaking.
In our chiral SU(3) quark model, one-gluon-exchange (OGE) governs 
the short range and scalar chiral field as well as pseudoscalar chiral field 
are induced for restoring the chiral symmetry. As is well known, the short range mechanism
of the quark-quark interaction  mechanism is still an open problem, and
people is debating whether the OGE plays a dominating role
in the short range, or vector meson exchange does, or both of them are important.
Thus to examine the short range mechanism,
we developed our chiral SU(3) quark model into the extended chiral SU(3) quark model 
in which the vector meson exchange is included.
During the past few years,
both the chiral SU(3) quark model and the extended chiral SU(3) quark model have
appeared to be quite successful in
reproducing the spectra of the baryon ground states, the binding
energy of deuteron, the nucleon-nucleon ($NN$), kaon-nucleon ($KN$)
scattering phase shifts, and the hyperon-nucleon ($YN$) cross
sections \cite{zhang1, zhang2, dai, fhuang04kn, fhuang04nkdk}.

Recently, the chiral quark model was also employed to study the
interactions and structures of the heavy-quark systems by using the
Resonating Group Method (RGM) \cite{liu1, liu2, wangwlsigc}.  Here
we will use the same chiral quark model to study the $D\bar{D}$ and
$B\bar{B}$ interactions. Different from the RGM method, we have
derived analytical forms of the total interaction potentials between
the two S-wave heavy mesons as discussed in Refs. \cite{li,li1}.
We thoroughly investigate the
possible bound states of  $D\bar{D}$ and $B\bar{B}$ systems by
solving the Schr\"{o}dinger equation with our analytical potentials
between the two clusters, anticipating that this method would give a
more accurate description of the short-range interaction between the
two clusters than the RGM does.

The paper is organized as follows. In section \ref{sec:formulism},
the framework of our chiral quark model is briefly introduced, and
the analytical forms of the effective interaction potentials between
the two S-wave heavy mesons under our chiral quark model are given.
The bound state solutions for the $D\bar{D}$ and $B\bar{B}$ systems
are shown and discussed in Sec.~\ref{sec:result}. Finally, a short
summary is given in Sec.~\ref{sec:sum}.

\section{Formulation} \label{sec:formulism}

In our chiral SU(3) quark model, we consider the scalar
meson-exchange interaction and the pseudoscalar meson-exchange
interaction which are induced from quark-chiral field interaction,
OGE interaction and confinement potential.
Moreover, in our extended chiral SU(3) quark model, we also take the
vector meson-exchange interactions into account which would almost
replace the  OGE interaction in the original chiral SU(3) quark model \cite{dai}.
In principle, the axial vector meson exchange should also be included
to keep the chiral symmetry. However, as the axial mesons are much heavier than
the chiral symmetry breaking scale, they are simply omitted in our calculation.

The framework of our models has been discussed extensively in the literature
\cite{zhang1,zhang2,dai,fhuang04kn,fhuang04nkdk,liu1,liu2,wangwlsigc,
wang2007,wang2008,wang2010,wang2011}.
In this work, different from the RGM calculation,
 the internal kinetic energies and the internal interactions of each meson
 are not necessary to be calculated.
As a result, the Hamiltonian of relative motion in this work reads
\begin{eqnarray}
H=T_{rel}+V_{eff},
\end{eqnarray}
where $T_{rel}$ is the kinetic energy operator of the relative
motion between the two mesons, and $V_{eff}$ is the effective
interaction potential derived from the quark-quark (quark-antiquark)
interaction between two mesons by integrating the internal
coordinates
 $\vec{\xi}_1$ and $\vec{\xi}_2$ of two mesons:
\begin{eqnarray}
V_{eff}=\sum_{ij}\int\varphi_1^*(\vec{\xi}_1)
\varphi_2^*(\vec{\xi}_2)V(\vec{r}_{ij})\varphi_1(\vec{\xi}_1)
\varphi_2(\vec{\xi}_2) d\vec{\xi}_1d\vec{\xi}_2,
\end{eqnarray}
while $\varphi_1(\vec{\xi}_1)$ and $\varphi_2(\vec{\xi}_2)$ are the
intrinsic wavefunctions of two S-wave mesons, taken as one-Gaussian form:
\begin{eqnarray}
\varphi(\vec{\xi})=\big(\frac{\mu\omega}{\pi}\big)^{3/4}
e^{-\frac{\mu\omega}{2}\xi^2}.
\end{eqnarray}
Here, $\mu$ is the reduced mass of the two quarks inside each meson
and $\omega$ is the harmonic-oscillator frequency of the meson
intrinsic wavefunction. $V(\vec{r}_{ij})$ in Eq. 2 represents the
interactions between the $i$-th light quark or antiquark in the first
meson and the $j$-th light quark or antiquark in another.

 In this work, because there is no color-interrelated interaction between the two
color-singlet clusters, such as OGE interaction and
confinement potential, we only consider the meson-exchange
interactions. Therefore, for the chiral SU(3) quark model
\begin{eqnarray}
V(\vec{r}_{ij})=\sum^8_{a=0} V^{\sigma_a}(\vec{r}_{ij})+\sum^8_{a=0}
V^{\pi_a}(\vec{r}_{ij}),
\end{eqnarray}
and for the extended chiral SU(3) quark model
\begin{eqnarray}
V(\vec{r}_{ij})=\sum^8_{a=0} V^{\sigma_a}(\vec{r}_{ij})+\sum^8_{a=0}
V^{\pi_a}(\vec{r}_{ij})+\sum^8_{a=0} V^{\rho_a}(\vec{r}_{ij}),
\end{eqnarray}
with $V^{\sigma_a}(\vec{r}_{ij})$ and $V^{\pi_a}(\vec{r}_{ij})$
 being the interactions respectively induced
from scalar meson exchange and  pseudoscalar meson exchange.
$V^{\rho_a}(\vec{r}_{ij})$ indicates the vector meson exchange
 interaction. For quark-quark (antiquark-antiquark) interaction,
$V^{\sigma_a}(\vec{r}_{ij})$, $V^{\pi_a}(\vec{r}_{ij})$
and  $V^{\rho_a}(\vec{r}_{ij})$ have been described in detail in Refs.
\cite{zhang1,zhang2,dai,
fhuang04kn,fhuang04nkdk,liu1,liu2,wangwlsigc,wang2007,wang2008,wang2010,
wang2011,li1}:
\begin{eqnarray}
V^{\sigma_a}(\vec{r}_{ij})&=& -C(g_{ch},m_{\sigma_a},\Lambda) X_1(m_{\sigma_a},\Lambda,r_{ij})
\left(\lambda^a_i\lambda^a_j\right),\label{scalar} \\
V^{\pi_a}(\vec{r}_{ij})&=& C(g_{ch},m_{\pi_a},\Lambda) \frac{m^2_{\pi_a}}{12m_im_j}
X_2(m_{\pi_a},\Lambda,r_{ij})  \times \left(\sigma_i\cdot\sigma_j\right)
\left(\lambda^a_i\lambda^a_j\right), \label{pseudoscalar}\\
V^{\rho_a}(\vec{r}_{ij}) &=& C(g_{\rm chv},m_{\rho_a},\Lambda)\Bigg[X_1(m_{\rho_a},
\Lambda,r_{ij}) + \frac{m^2_{\rho_a}}{6m_im_j} \times \left(1+\frac{f_{\rm chv}}
{g_{\rm chv}} \frac{m_i+m_j}{M_N}+\frac{f^2_{chv}}{g^2_{chv}}
\frac{m_im_j}{M^2_N}\right)   \nonumber \\
&& \times \,  X_2(m_{\rho_a},\Lambda,r_{ij}) \, (\sigma_i\cdot\sigma_j)\Bigg]
\left(\lambda^a_i\lambda^a_j\right),\label{vector}
\end{eqnarray}
with
\begin{eqnarray}
C(g_{ch},m,\Lambda) &=& \frac{g^2_{ch}}{4\pi} \frac{\Lambda^2}{\Lambda^2-m^2} m, \\
\label{ev1} X_1(m,\Lambda,r_{ij}) &=& Y(mr_{ij})-\frac{\Lambda}{m} Y(\Lambda r_{ij}), \\
\label{ev2} X_2(m,\Lambda,r_{ij}) &=& Y(mr_{ij})-\left(\frac{\Lambda}{m}\right)^3 Y(\Lambda r_{ij}), \\
Y(x) &=& \frac{1}{x}e^{-x},
\end{eqnarray}
where $\lambda^a$ is the Gell-Mann matrix in flavor space, and $\Lambda$ is
the cutoff mass which indicates the chiral symmetry breaking scale. 
$m_i$ and $m_j$  are the masses of the $i$-th light quark or antiquark in the first
meson and the $j$-th light quark or antiquark in another respectively,
 while $m_{\sigma_a}$, $m_{\pi_a}$
and $m_{\rho_a}$  in Eqs. \ref{scalar}, \ref{pseudoscalar},
and \ref{vector} are the masses of the scalar nonets, the
pseudoscalar nonets and the vector nonets, respectively. $M_N$ in
Eq. \ref{vector} is a mass scale usually taken as the mass of
nucleon \cite{dai}. $g_{ch}$ is the coupling constants for the
scalar and pseudoscalar nonets. $g_{chv}$ and $f_{chv}$ are the
coupling constants for the vector coupling and tensor coupling of
vector nonets, respectively.

In $D\bar{D}$ and $B\bar{B}$ systems, we don't consider
the one-meson exchange interactions between two heavy quarks or
between one heavy quark and one light quark,
because if one wants to include the interactions related to heavy quarks,
the heavy meson-exchanges as well as light meson exchanges must be also considered
simultaneously, and these interactions are beyond our SU(3) models. By using the method
described in Refs.  \cite{li,li1} and integrating the internal
coordinates of two mesons, we get the analytical effective
interaction potentials between two mesons $D\bar{D}$($B\bar{B}$) as
\begin{eqnarray}
V_{eff}(\vec{R})=\sum^8_{a=0} V_{q\bar{q}}^{\sigma_a}(\vec{R})+\sum^8_{a=0}
V_{q\bar{q}}^{\pi_a}(\vec{R})+\sum^8_{a=0} V_{q\bar{q}}^{\rho_a}(\vec{R}), \nonumber
\end{eqnarray}
with
\begin{eqnarray}
V_{q\bar{q}}^{\sigma_a}(\vec{R})&=& -G_{\sigma_a}C(g_{ch},m_{\sigma_a},\Lambda)
X_{1q\bar{q}}(m_{\sigma_a},\Lambda,R)\left(\lambda^a_q\lambda^a_{\bar{q}}
\right),\label{eff:scalar} \\
V_{q\bar{q}}^{\pi_a}(\vec{R})&=& G_{\pi_a}C(g_{ch},m_{\pi_a},\Lambda)
\frac{m^2_{\pi_a}}{12m_qm_{\bar{q}}} X_{2q\bar{q}}(m_{\pi_a},\Lambda,R)
 \times \left(\sigma_q\cdot\sigma_{\bar{q}}\right)\left(\lambda^a_q
\lambda^a_{\bar{q}}\right), \label{eff:pseudoscalar}\\
V_{q\bar{q}}^{\rho_a}(\vec{R}) &=& G_{\rho_a}C(g_{chv},m_{\rho_a},\Lambda)
\Bigg[X_{1q\bar{q}}(m_{\rho_a},\Lambda,R)
+ \frac{m^2_{\rho_a}}{6m_qm_{\bar{q}}}\left(1+\frac{f_{chv}}{g_{chv}}
\frac{m_q+m_{\bar{q}}}{M_N}+\frac{f^2_{chv}}{g^2_{chv}}
\frac{m_qm_{\bar{q}}}{M^2_N}\right)   \nonumber \\
&& \times X_{2q\bar{q}}(m_{\rho_a},\Lambda,R) \, (\sigma_q\cdot
\sigma_{\bar{q}})\Bigg] \left(\lambda^a_q\lambda^a_{\bar{q}}\right).
\label{eff:vector}
\end{eqnarray}
Here, $G_{\sigma_a,\pi_a,\rho_a}$ is the $G$-parity of the
exchanged meson, and $\vec{R}$ is the relative coordinate between two
different mesons, namely, the relative coordinate between the two
centers-of-mass coordinates of the two mesons, and
\begin{eqnarray}
\label{eff:ev1} X_{1q\bar{q}}(m,\Lambda,R) &=& Y_{q\bar{q}}(mR)
-\frac{\Lambda}{m} Y_{q\bar{q}}(\Lambda R), \\
\label{eff:ev2} X_{2q\bar{q}}(m,\Lambda,R) &=& Y_{q\bar{q}}(mR)
-\left(\frac{\Lambda}{m}\right)^3 Y_{q\bar{q}}(\Lambda R).
\end{eqnarray}
In above equations, $m_q$ and $m_{\bar{q}}$ are masses of the light
quark and antiquark, respectively.
The modified Yukawa term in Eqs. \ref{eff:ev1} and \ref{eff:ev2} reads
\begin{eqnarray}
Y_{q\bar{q}}(mR)&=&\frac{1}{2mR}e^{\frac{m^2}{4\beta}}\bigg\{e^{-mR}
\Big\{1-erf\Big[-\sqrt{\beta}(R-\frac{m}{2\beta})\Big]\Big\}
-e^{mR}\Big\{1-erf\Big[\sqrt{\beta}(R+\frac{m}{2\beta})\Big]\Big\}\bigg\}.
\label{modified Yukawa}
\end{eqnarray}
Here,
\begin{eqnarray}
\beta&=&\frac{\mu_{q\bar{Q}}\mu_{Q\bar{q}}\omega}{\mu_{q\bar{Q}}
\left(\frac{m_{Q}}{m_Q+m_{\bar{q}}} \right)^2+\mu_{Q\bar{q}}
\left(\frac{m_{\bar{Q}}}{m_q+m_{\bar{Q}}} \right)^2}.
\end{eqnarray}
 $m_Q$ and $m_{\bar{Q}}$ are masses of the heavy quark and
antiquark respectively, and $\mu_{qQ}=\frac{m_qm_Q}{m_q+m_Q}$.

There are some necessary parameters in the potentials of our chiral
quark model. In this work, we adopt the parameters determined in our
previous works
\cite{dai,liu1,liu2,zhang1,fhuang04kn,fhuang04nkdk,zhang2,wangwlsigc,wang2007,
wang2008, wang2010}. The up/down quark mass $m_q$ is 
fitted as the nucleon mass and taken as $M_N/3\sim 313$ MeV. The
coupling constant for the scalar and pseudoscalar chiral fields
$g_{ch}=2.621$ is fixed by the relation of
\begin{eqnarray}
\frac{g^{2}_{ch}}{4\pi} =\frac{9}{25} \frac{g^{2}_{NN\pi}}{4\pi}
\frac{m^{2}_{u}}{M^{2}_{N}},\nonumber
\end{eqnarray}
with $g^{2}_{NN\pi}/4\pi=13.67$ determined from experiments.

In our extended chiral SU(3) quark model, the vector
coupling constant $g_{chv}$ and tensor coupling constant $f_{chv}$ in Eqs. \ref{vector},
\ref{eff:vector} are fitted by the mass difference between $N$ and $\Delta$,
when the strength of the OGE is taken to be almost zero.
When the tensor coupling is neglected, $g_{chv}=2.351$ and $f_{chv}=0$;
when the tensor coupling is considered, $g_{chv}=1.973$ and $f_{chv}=1.315$.
The harmonic-oscillator frequency $\omega$, equal to $1/(m_u b_u^2)=1/(m_c b_c^2)$
where $b_u$ is fitted by the $N-N$ scattering phase shifts, is taken as $2.522 fm^{-1}$ in the
chiral SU(3) quark model and $3.113 fm^{-1}$ in the extended chiral
SU(3) quark model. In our calculation, the masses of the mesons are
taken from the PDG \cite{PDG2010}, except the $\sigma$ meson, which
does not have a well-defined value. Here $m_\sigma$ is obtained by
fitting the binding energy of the deuteron \cite{dai}. It is
$m_\sigma=595$ MeV in our chiral SU(3) quark model, 535
MeV  for neglecting tensor coupling  and 547  MeV for
considering tensor coupling in our extended chiral SU(3) quark
model. The cutoff mass $\Lambda$ is the chiral symmetry
breaking scale and taken as $1100$  MeV as a convention.

The remaining parameters to be determined are the heavy quark masses $m_c$
and $m_b$. In our work, we find that the final results are not sensitive
to the variation of the heavy quark masses, and we take  $m_c=1430$  MeV
\cite{zhanghx1} and $m_b=4720$  MeV \cite{zhanghx2} as typical values.

\section{NUMERICAL SOLUTIONS}\label{sec:result}

\subsection{\textbf{$D\bar{D}$}}

\begin{figure}
  \centering
  \subfigure[]{
    \label{DD} 
    \includegraphics[width=3in, width=7.5cm]{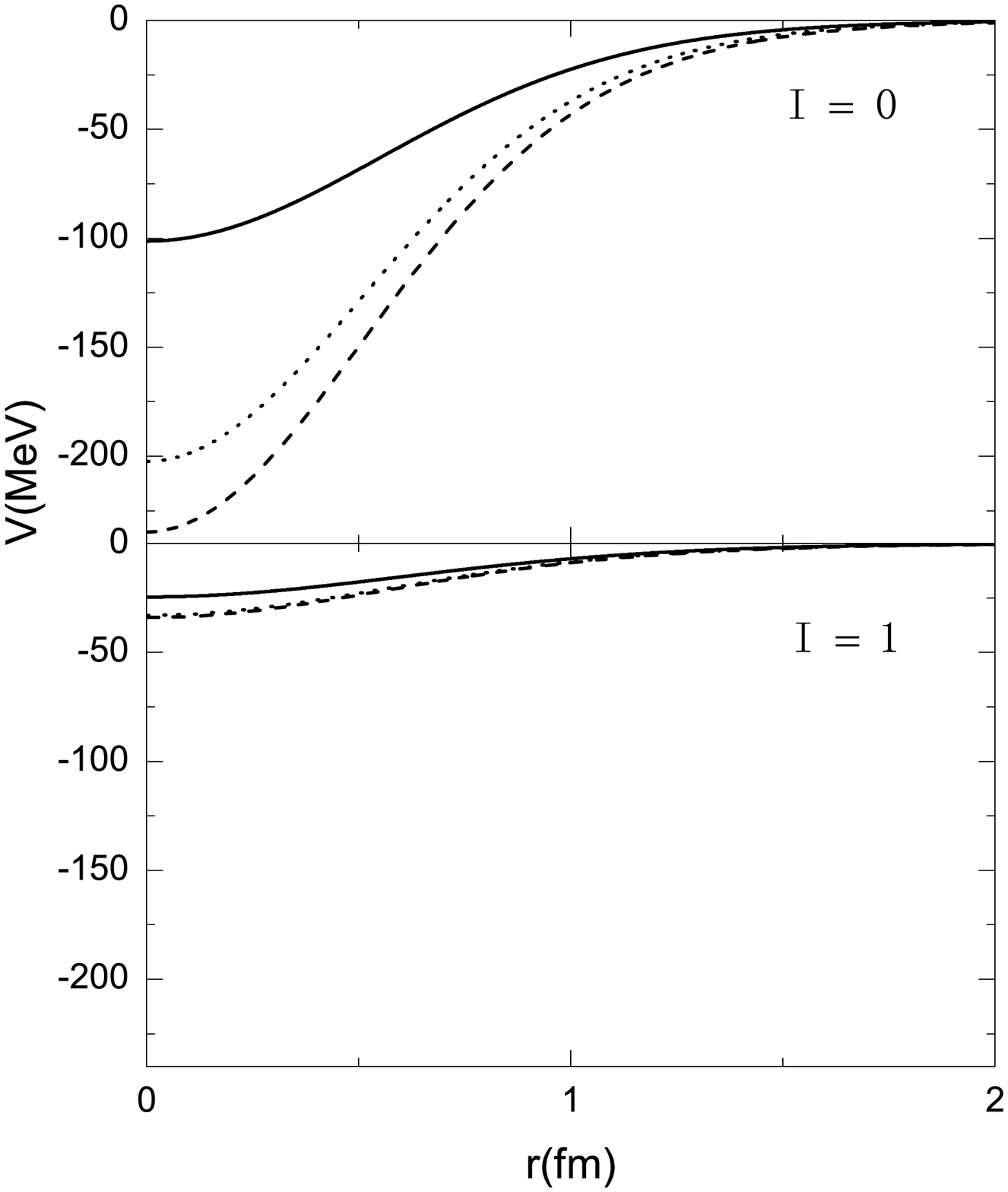}}
  \hspace{0.5in}
  \subfigure[]{
    \label{BB} 
    \includegraphics[width=3in,width=7.5cm]{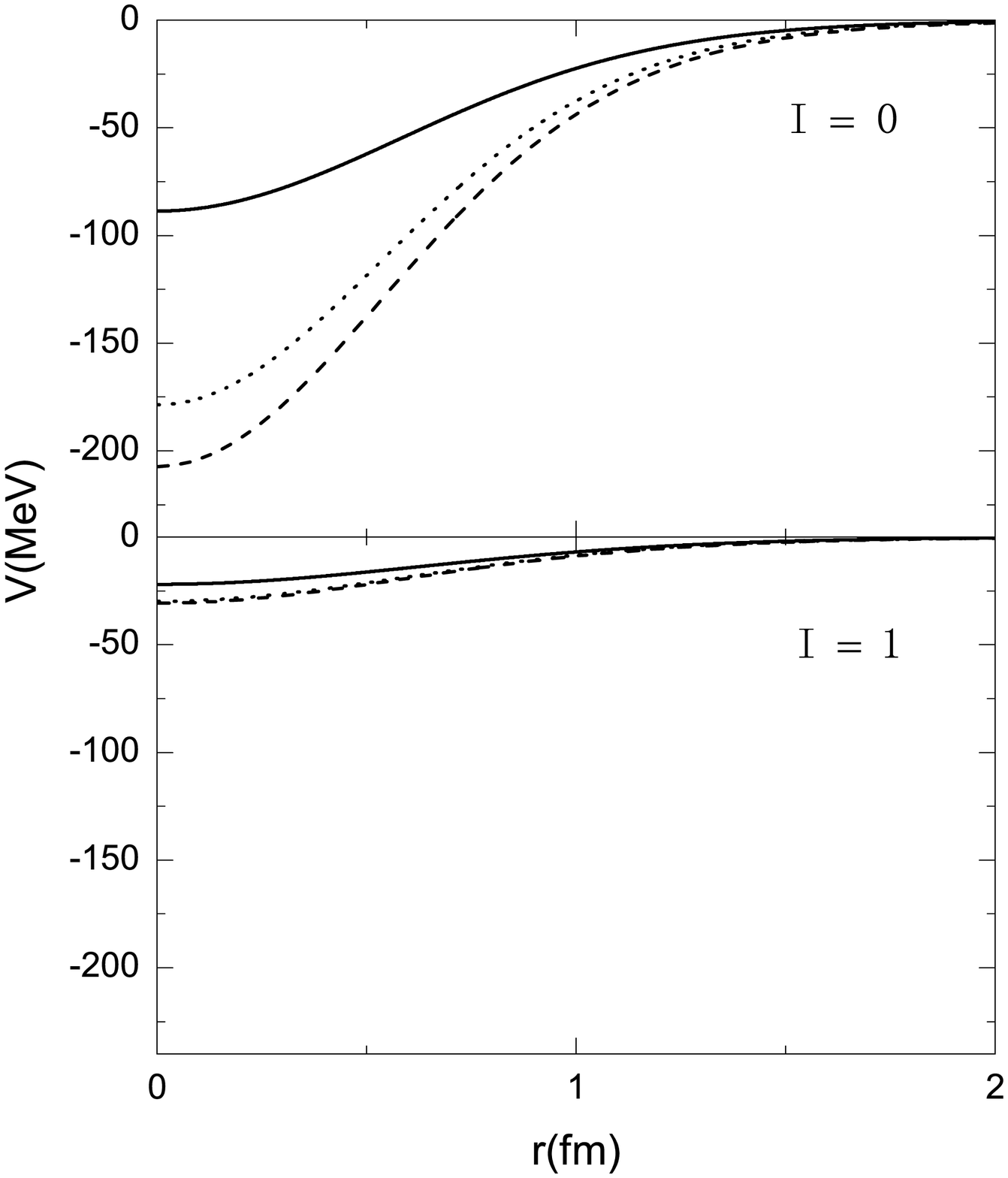}}
  \caption{The total interaction potential of  $D\bar{D}$ and $B\bar{B}$
systems. \ref{DD} the total interaction of $D\bar{D}$ system and
\ref{BB} the total interaction of $B\bar{B}$ system. The solid,
dashed and dotted lines represent  the results obtained from the
chiral SU(3) quark model, the extended chiral SU(3) quark model
without and with the tensor coupling of vector field,
respectively.}
  \label{fig:subfig} 
\end{figure}

Here we study the $D\bar{D}$ system with different isospin $I$.
Following the approach introduced in section II, we get the
analytical effective interactions between $D\bar{D}$ in our chiral
quark model which are depicted in Fig. \ref{DD}. From Fig. \ref{DD},
one sees that the $D\bar{D}$ interaction is attractive, especially
for isospin $I=0$ case. To study if such an attraction is strong
enough to bind the $D\bar{D}$ system, we solve the Schr\"{o}dinger
equation with the programs developed in Refs. \cite{schoberl1985,
lucha1999}, and list the obtained binding energies in Table
\ref{DDt}. Because the mass difference between
$D\bar{D}$ and $D^*\bar{D}^*$ is not very large, we also consider the coupled effect of
$D^*\bar{D}^*$ on $D\bar{D}$. We carry out a perturbative calculation to see the
contributions of the off-diagonal elements of this coupled channel
to the binding energies of $D\bar{D}$ and list them in Table
\ref{DDt}. From Table \ref{DDt}, we see there is only one
$I^G(J^{PC})=0^{+}(0^{++})$ $S$-wave $D\bar{D}$ bound state with a
binding energy 3--35 MeV in our chiral quark model.

{\small
\begin{table}[htbp] \caption{The binding energy and the root of
mean square radius of $D\bar{D}$ binding system. The binding energy is listed in such a
way: B+$\Delta$B, B is the binding energy deduced in the single channel calculation and $\Delta$B is the perturbation
correction value deduced from the off-diagonal elements of the coupled channel of $D\bar{D}$
and $D^*\bar{D}^*$.}\label{DDt}
\begin{tabular*}{140mm}{@{\extracolsep\fill}lcccccc}
\hline\hline
 & \multicolumn{2}{c}{$\chi$-SU(3) QM } & \multicolumn{4}{c}{Ex. $\chi$-SU(3) QM }\\
\cline{4-7}
 &                       &             &\multicolumn{2}{c}{$g_{chv}=2.351$,$f_{chv}=0$} &\multicolumn{2}{c}{$g_{chv}=1.973$,$f_{chv}=1.315$}  \\
\hline
 & B+$\Delta$B(MeV) &$r_{rms}$(fm)             &B+$\Delta$B(MeV)         &$r_{rms}$(fm)                 &B+$\Delta$B(MeV)                 &$r_{rms}$(fm)\\
\hline
$I=0$& 1.0+2.3         &3.7        & 33.3+1.4         &1.0                        & 21.6+0.2                  &1.1 \\
$I=1$& $-$              &$-$           & $-$               &$-$                         & $-$                        &$-$\\
\hline\hline
\end{tabular*}
\end{table}}

Further analysis shows that the $D\bar{D}$ interaction is dominated
by $\sigma$, $\sigma'$, $\omega$ and $\rho$ exchanges. For the $I$=0
case, in the chiral SU(3) quark model, both $\sigma$ and $\sigma'$
exchanges provide attractive interactions, so the total interaction
is strong enough to form a $D\bar{D}$ bound state. In the extended
chiral SU(3) quark model, the contributions of vector meson exchange
are also included, and $\rho$ and $\omega$ exchanges provide
additional attraction. Therefore, the $D\bar{D}$ system has a larger
binding energy as shown in Table \ref{DDt}. We see that
the perturbative contribution of the coupled channel is not very
large compared to the results of the single channel and wouldn't
obviously change the main feature of the binding solution. The radii
of $D$ and $\bar{D}$ both are about $0.54 fm$ in the
chiral SU(3) quark model and $0.49 fm$ in the extended chiral SU(3) quark model
(since $\omega$ is different in the two models), and we find that
the rms radius of this $0^{+}(0^{++})$ $D\bar{D}$ bound state is
bigger than the sum of the radii of $D$ and $\bar{D}$, so we
conclude that this $D\bar{D}$ could form a molecule and this
$0^{+}(0^{++})$ $D\bar{D}$ molecular state has a mass
3695--3726 MeV. For the $I=1$ case, in the chiral
SU(3) quark model, $\sigma$ exchange provides attraction but
$\sigma'$ exchange provides repulsion, thus the total attractive
interaction is too weak to make a $D\bar{D}$ bound state. In the
extended chiral SU(3) quark model, the additional attraction
provided by $\omega$ exchange and the additional repulsion provided
by $\rho$ exchange almost cancel each other, as a result, the total
interaction of all the meson exchanges is still too weak to bind
$D\bar{D}$. Thus, no $1^{-}(0^{++})$ $D\bar{D}$ bound state can be
obtained due to the insufficiency of the attraction.

\subsection{\textbf{$B\bar{B}$}}

{\small
\begin{table}[htbp] \caption{The binding energy  and the root of mean square radius of $B\bar{B}$ binding system. The binding energy is listed in such a way: B+$\Delta$B, B is the binding energy deduced in the single channel calculation and $\Delta$B is the perturbation correction value deduced from the off-diagonal elements of the coupled channel of $B\bar{B}$ and $B^*\bar{B}^*$.}\label{BBt}
\begin{tabular*}{140mm}{@{\extracolsep\fill}lcccccc}
\hline\hline
 & \multicolumn{2}{c}{$\chi$-SU(3) QM } & \multicolumn{4}{c}{Ex. $\chi$-SU(3) QM }\\
\cline{4-7} &                       &             &\multicolumn{2}{c}{$g_{chv}=2.351$,$f_{chv}=0$} &\multicolumn{2}{c}{$g_{chv}=1.973$,$f_{chv}=1.315$}  \\
\hline
 & B+$\Delta$B(MeV) &$r_{rms}$(fm)             &B+$\Delta$B(MeV)         &$r_{rms}$(fm)                 &B+$\Delta$B(MeV)                 &$r_{rms}$(fm)\\
\hline
$I=0$& 19.2+1.2        &0.9           & 81.8+11.3         &0.6                         & 64+2.6              &0.6 \\
$I=1$& $-$             &$-$           & $-$               &$-$                         & $-$                 &$-$\\
\hline\hline
\end{tabular*}
\end{table}}

Considering the resemblance between $D\bar{D}$ and $B\bar{B}$, the
one meson exchange interaction potentials should have similar
properties. In our calculation we find the total interaction
potential of $B\bar{B}$ system is also attractive, as shown in Fig.
\ref{BB}. By Solving the Schr\"{o}dinger equation and
considering the perturbative correction of the coupled channel of
$B^*\bar{B}^*$ to the binding energies of $B\bar{B}$, we list our
results in Table \ref{BBt}. We notice our results are
similar to the ones of Liu {\it et al.} \cite{liu1}.
It should be emphasized that the $B\bar{B}$ system is easier to form
a bound state than the $D\bar{D}$ system, because $B\bar{B}$ system
has a much heavier reduced mass 2640 MeV than that of $D\bar{D}$
system of 932 MeV. We also find that the radii of $B$
and $\bar{B}$ are both about $0.52 fm$ in the chiral SU(3) quark model and
$0.46 fm$ in the extended chiral SU(3) quark model. However, according to
Table \ref{BBt} we see that the rms radius of this $B\bar{B}$ bound
state is smaller than the sum of the radii of $B$ and $\bar{B}$.
This feature means $B$ and $\bar{B}$ are overlapped with each other,
and this $0^{+}(0^{++})$ $B\bar{B}$ bound state
seems not a molecular state but might be a tetra-quark state
with a mass 10467-10540 MeV in our chiral quark models.

\section{summary}\label{sec:sum}

In this work, we have studied the bound state problem of $D\bar{D}$
and $B\bar{B}$ systems with an analytical effective interaction
potential in our chiral quark model. We find only one
$0^+(0^{++})$ $D\bar{D}$ molecule could exist with
a mass 3695--3726 MeV, agreeing with the
predictions of Refs. \cite{wong,huang,valcarce1,valcarce2,
valcarce3,liux1,liux2,liux3,liu1,ke}. For the $0^+(0^{++})$
$B\bar{B}$ system, they are bound even deeper than the $D\bar{D}$,
however, the small rms radii mean the overlap is sizeable and, therefore,
it may not be interpreted as molecular states, but a tetra-quark state.

We expect future experimental measurements would test
the results of our chiral SU(3) quark model and the extended chiral SU(3) quark model.
Since our calculation shows the important role of the vector meson exchange,
we hope that the future measurement could recognize whether vector meson-exchange
or one-gluon-exchange dominates the short range interaction between quarks.

\begin{acknowledgments}
This project was supported£¬ in part£¬ by National Natural Science
Foundation of China (Grant Nos. 11105158, 11035006, 1097514610,
11103500632, and 11261130311), by the DFG and the NSFC through funds
provided to the Sino-Germen CRC 110 ``Symmetries and the Emergence
of Structure in QCD'',  by the Ministry of Science and Technology of
China (Grant No. 2009CB825200),  and by China Postdoctoral Science
Foundation (Grant No. 20100480468).
\end{acknowledgments}

\end{document}